\documentclass[12pt,preprint]{aastex}
\usepackage[usenames]{color}

\def\l{$\lambda$}

\def\rfe{$R_{\rm FeII}$}

\def\ltsima{$\; \buildrel < \over \sim \;$}
\def\ltsim{\lower.5ex\hbox{\ltsima}}
\def\gtsima{$\; \buildrel > \over \sim \;$}

\def\gtsim{\lower.5ex\hbox{\gtsima}}

\def\cm3{cm$^{-3}$\/}
\def\hb{{\sc{H}}$\beta$\/}

\def\o4363{{\sc{[Oiii]}}$\lambda$4363\/}

\def\fe{{\sc{Fe}}\/}

\def\vr{{$v_{\mathrm r}$}}

\def\fe76087{{\sc [Fe vii]}$\lambda$6087\/}
\def\oiii{{\sc [Oiii]}$\lambda$5007}

\def\kms{km~s$^{-1}$}

\slugcomment{}

\shorttitle{No FeII Redshift in Quasars}
\shortauthors{Sulentic et al.}

\begin{document}
\title{NO EVIDENCE FOR A SYSTEMATIC FEII EMISSION LINE REDSHIFT IN TYPE 1 AGN
}
\author{Jack W. Sulentic%\altaffilmark{1}
}
\affil{Instituto de Astrof\'isica de Andaluc\'ia, CSIC, Spain}
\email{sulentic@iaa.es}
\author{Paola Marziani}
\affil{INAF, Astronomical Observatory of Padova, Italy}
\email{paola.marziani@oapd.inaf.it}
\author{Sebastian Zamfir}
\affil{University of Wisconsin, Stevens Point, USA}
\email{szamfir@uwsp.edu}
\and
\author{Zachary A. Meadows}
\affil{University of Wisconsin, Stevens Point, USA}
\email{Zachary.A.Meadows@uwsp.edu}

\begin{abstract}
We test the recent claim by Hu et al. (2008) that FeII emission in Type 1 AGN shows a
systematic redshift relative to the local source rest frame and broad-line H$\beta$. We compile high s/n median composites using SDSS spectra from both the Hu et al. sample and our own sample of the 469 brightest DR5 spectra. Our composites are generated in bins of FWHM H$\beta$ and FeII strength as defined in our 4D Eigenvector 1 (4DE1) formalism. We find no evidence for a systematic FeII redshift and consistency with previous assumptions that FeII shift and width (FWHM) follow H$\beta$ shift and FWHM in virtually all sources. This result is consistent with the hypothesis that FeII emission (quasi-ubiquitous in type 1 sources) arises from a broad-line region with geometry and kinematics the same as that producing the Balmer lines.
\end{abstract}

\section{Introduction}

Spectra of Type 1 AGN show a diversity of broad and narrow emission lines
that provide direct insights into the structure and kinematics of
photoionized, and otherwise excited, gas in the vicinity of the putative
central massive object. Broad emission lines, like much studied H$\beta$
\citep[e.g.,][hereafter Z10]{zamfiretal10}, are thought to arise in or near an accretion disk
acting as the fuel reservoir for the central supermassive black hole
(log M$\sim 7-9.5$ M$_{\odot}$). H$\beta$ shows a diversity of line widths
as well as profile shifts and asymmetries \citep{sulenticetal00a}. Despite
this diversity some systematics have emerged and are best highlighted via
the concept of two Type 1 AGN populations \citep{sulenticetal00b}.

Population A show the smallest broad-line widths FWHM H$\beta$=1000-4000 \kms and includes the Narrow line Seyfert 1 (NLSy1) sources (FWHM $<$ 2000 \kms). Pop. A H$\beta$ profiles are currently best fit by a single Lorentz function.
Population B sources show FWHM H$\beta$=4000-12000 \kms and require
two Gaussians (one unshifted and one redshifted) for a reasonable profile
description. ``Broad-line'' H$\beta$ profiles as narrow as FWHM =
500 \kms \citep{zhouetal06} and as broad as FWHM = 40000 \kms \citep{wangetal05}
have been found. Pop. A is predominantly radio-quiet while pop. B
involves a mix of radio-quiet and the majority of radio-loud quasars.

Broad- and narrow-line profile shifts are known and the phenomenology can
be confusing. Narrow emission lines like [OIII]\l5007{\AA} are regarded as a
reliable measure of the local quasar rest frame except in the case of ``blue
outliers'', usually found in sources with FWHM H$\beta$= 1500-3500 \kms
and weak [OIII] \citep{veroncettyetal01,zamanovetal02, marzianietal03b,boroson05}.
Blue outliers show [OIII] blueshifts as large as $\sim$1000 \kms. No
pop. B sources with blueshifted [OIII] are known at low z (or luminosity). Careful use of [OIII] and H$\beta$ narrow line as rest frame measures suggests that broad H$\beta$ in pop. A sources rarely shows a systematic red or  blue
shift above the FWHM profile level. A blueshifted component or asymmetry is observed in some extreme FeII strong pop. A sources \citep{Marzianietal10}. Pop. B sources show more complex line shift properties. The H$\beta$ profile usually
shows two components: 1) a ``classical'' broad component (BC; FWHM = 4000 -- 5000 \kms)
with zero or small (red or blue) shift, and 2) a very broad
(VBC; 10000 \kms) and redshifted ($\gtsim$1000 \kms) component. Composites involving
the 469 brightest SDSS-DR5 quasars\footnote{We acknowledge the use of SDSS quasar spectra: http://www.sdss.org/collaboration/credits.html} suggest that these two components represent the
underlying stable structure of H$\beta$ in pop. B sources.

Broad FeII emission has been found in type 1 quasars since the era of photographic spectroscopy
in the '60s. FeII emission blends are almost ubiquitous in a sample of the brightest (usually
highest s/n) SDSS quasars (Z10). Circumstantial evidence has
accumulated supporting the assumption that FeII emission arises in or near the emitting clouds that
produce other low ionization lines like H$\beta$ (see e.g., \citealt{borosongreen92, marzianietal03a, sulenticetal06}). FWHM FeII appears to correlate with FWHM H$\beta$\ over the full range where FeII
can be detected (FWHM=1000-12000 \kms). This can be clearly seen at low $z$\ by observing the shape
(e.g., smoothness) of the FeII 4450-4700 {\AA}\ blue blend (and the FeII multiplet 42 line at 5018 {\AA})
near [OIII]\l5007{\AA}. In pop. A sources the blend resolves into individual lines
while it becomes much smoother in pop. B sources. Sources with the strongest FeII emission also
show a weakening of H$\beta$ emission as expected if the latter is collisionally quenched in the
same dense medium where strong FeII emission can be produced \citep{gaskell85}.

\section{A Systematic FeII Redshift?}

Obviously systematic line shifts place important constraints on models for the geometry and kinematics
of the broad line region. The most famous example involves a systematic blueshift of high ionization
lines (e.g., CIV \l1549{\AA})  relative to low ionization lines (e.g., Balmer) especially in pop. A sources \citep[e.g.,][]{gaskell82,sulenticetal07,richardsetal11,wangetal11}. Evidence was recently advanced \citep[][hereafter H08]{huetal08}  for the existence of a {\em systematic} redshift of FeII relative to
[OIII]\l5007 (and hence the Balmer lines) in a majority of Type 1 quasars. This result, along with a
narrower estimated FeII line width, has been ascribed to FeII emission arising in a region with dynamics
dominated by infall and located at larger radius than the region producing the bulk of H$\beta$.
H08 argue that the amplitude of the shifts correlates inversely with source Eddington ratio (L/L$_{Edd}$$\equiv$$\eta$). Interpretations
for such an FeII redshift have already appeared \citep{ferlandetal09,shieldsetal10,boroson11} reflecting
the potential importance of such a first-order kinematic signature. Having worked on line spectra and profile shifts for many  years we were surprised by the H08 claims and decided to test the hypothesis of a
systematic FeII redshift. Could we have missed it?

First let us consider what we know. Most pop. A quasars show relatively symmetric unshifted Lorentz-like
H$\beta$ profiles with FWHM$<$4000 \kms. In our work using the brightest ($g <$ 17.5 or $i <$ 17.5; \citealt{zamfiretal08}) SDSS DR5 quasars we processed spectra for $\sim$260 pop. A sources (from a sample of 469 quasars; Z10) and we found no evidence for a systematic shift of FeII lines relative to H$\beta$ or \oiii. Such an FeII shift should be easiest to detect in the brightest pop. A SDSS spectra with narrowest broad-line profiles and strongest FeII emission. It is immediately suspicious that more and larger FeII redshifts are claimed for pop. B sources. In only one pop. A source in our sample SDSS J0946+0139 do we find a
large H$\beta$ peak (90$\%$ intensity level) redshift of 1830 \kms. This source
is similar to OQ208  (\citealt{Marzianietal93} and discussed in H08) which shows $\Delta $ \vr$>$2000 \kms.
SDSS J0946 is the only pop. A source with a large FeII redshift in our Z10 sample (1/260).
Z10 found 19 quasars with an H$\beta$ peak (9/10 fractional intensity) blueshifted more than -320 \kms and 4 sources with the peak redshift more than +320 \kms.
The remaining 241 pop. A sources showed no significant H$\beta$\ peak shift (figure 8 of Z10).
Best FeII template fits to these sources show no significant difference in
centroid redshift between FeII and H$\beta$. There are two possible causes of small and spurious
H$\beta$ (or FeII) shifts: 1) host galaxy contamination and 2) blue outliers.Except in rare cases host galaxy
contamination is unlikely to induce systematic redshifts with the amplitudes reported by H08.

Extreme blue outliers with [OIII] blueshifts in the
range 400-1000 \kms are rare and therefore cannot be the cause of the large and systematic
shifts reported in H08. In fact H08 selection criteria rejected sources likely to be  seriously affected
by 1) or 2).

H08 chose 4000+ sources from SDSS DR5 with computed s/n $\geq$ 10. Z10 also used DR5 where $\sim$94\% of
sources show s/n $\geq$ 10. Our sample was magnitude-limited with a slightly shallower redshift upper
limit (z=0.7 instead of 0.8). Why do we reach different conclusions about FeII shifts? A big part of
the answer could involve how s/n was computed. H08 compute s/n over the range 4430--5500$\rm{\AA}$.
This procedure overestimates the quality of the data because it includes major emission lines in the
computation. We compute s/n in the range 5600-5800$\rm{\AA}$, which is free of strong lines and
represents as close as one can approach to an estimate of continuum s/n near H$\beta$. Using our range
the H08 sample shows mean and median s/n values of 10.6 and 7.4, respectively; approx. 65\% show s/n $<$ 10. We find that only 182 spectra of our bright sample are included in the H08's. The majority of the H08's sources are lower s/n than those in our sample.

\section{{\sc specfit} Analysis}

\subsection{A2 and B1 Composite Spectra}

One cannot estimate reliable FeII line shifts using individual SDSS spectra for sources
fainter than about g$\sim$17-17.5. In rough order of importance our studies indicate that
the accuracy of FeII shift measures depends on: 1) FeII strength and FeII/H$\beta$\ profile widths,
2) spectral s/n and 3) if estimates depend heavily on fits to the  4430-4680 \AA\ blend,
strength of HeII 4686 emission. Typical individual spectra used by H08 {\em show} too low s/n to allow
convincing conclusions about FeII shift and width -- typical parameter uncertainties for
individual sources are much larger than the ones connected with our high s/n composites
(for a typical A2 source with s/n$\approx$20 uncertainties of shift estimates are larger than $\pm$1500\kms). Individual source spectra with large quoted FeII redshift and s/n near the
sample median were extracted  from the H08 sample and {\sc specfit } modelled. Using an FeII
template with fixed shifts ranging from zero up to the largest values quoted by H08, $\chi^2$
cannot distinguish between zero and e.g., 1000 \kms redshift in the majority of the sources.

The best recourse is to generate high s/n composite spectra. H08 argue that one cannot confirm
or refute the existence of a systematic FeII redshift using composite spectra because of the
large dispersion of FWHM, shifts and flux values for both H$\beta$ and FeII. This is likely true
for composites generated from random subsamples of sources but not true if one generates composites
over more limited ranges of parameter values. One can generate binned composites over limited
ranges of FWHM H$\beta$ and FeII strength using the 4DE1 formalism  (\citealt{sulenticetal02,bachevetal04,sulenticetal07}; Z10). 4DE1 bins A2 (FWHM H$\beta$= 1000-4000 \kms,
0.5 $\le$ \rfe $\le$ 1.0) and B1 (FWHM H$\beta$=4000 -- 8000 \kms, \rfe $\le$ 0.5) are  of
particular interest because they include the largest numbers of sources in random samples. {\sc specfit} analysis (\citealt{kriss94}; details in \citealt{marzianietal09}) of an A2 median composite involving n = 130 Z10 sources (s/n~90) gives a best-fit consistent with zero FeII redshift. The situation for the B1 composite (n=131 sources from Z10; s/n~110) is less constraining because lines are broader and FeII weaker. Table 1 reports FeII template shifts and 2$\sigma$\ uncertainties for {\sc specfit} tests discussed in  this paper. We also report peak shifts of \hb\ BC extracted from the best {\sc specfit} solutions  along with \hb\ ``core'' shifts measured at the centroid of the line peak after \hb\ NC subtraction. In no case do we find a significant shift between FeII and the rest frame or between FeII and \hb.  We also do not find any significant FeII shifts if we restrict to sources with L/L$_{Edd}$ ratio $\eta\le 0.1$\ (H08 suggested the shifts might be largest for low L/L$_{Edd}$ sources).

\subsection{A2 Composites Involving Only Sources with Large FeII Shifts?}

Since we find no evidence for systematic FeII redshifts in our Z10 bright quasar sample
composites it is useful to generate FeII shift composites using the H08 sample. We
generate them within the 4DE1 context thereby restricting the ranges of FWHM H$\beta$, FeII
relative strength (and likely also FWHM FeII) for each composite. Since the
distribution of FeII shifts shown in H08 is continuous we focus on the sources with largest
quoted shift values. If these shifts are not confirmed then smaller shifts are even
less likely to be real. We therefore focus on constructing median composites for all H08 sources falling in 4DE1 bins A2 and B1 with H08 FeII redshift estimates $\ge 1000$ \kms\ (figure 1).
Two composites were constructed for each spectra bin: 1) one with no restriction on FeII width
(H08 do not constrain FWHM FeII in their template fits so it is sometimes very different from
FWHM H$\beta$ ) and 2) one with FeII width constrained to the FWHM range of H$\beta$\ in a particular
bin (i.e., $\le 4000 $\kms\ for A2 and $ 4000 - 8000 $ \kms\ for B1). Upper and lower panels of figure 1 show  bins A2 (FWHM H$\beta$ $<$ 4000 \kms) and B1 (FWHM H$\beta$=4000-8000 \kms), respectively (n=156 for bin A2 and n=240 for bin B1). The s/n $\sim$ 55-60 for both composites. Spectra show best-fit {\sc specfit} models
superimposed. The left and center panels involved FeII templates fixed to the best fit and 1500 \kms
shifts, respectively. Our template prescription is described in \citet{marzianietal09}

Graphical results for the best-fits are shown  in the right panels  of figure 1.
Fits were performed over the range  $\approx$ 4470 -- 5450 \AA, where FeII and
continuum emission account for the total flux making it the safest region
for normalized $\chi^{2}$\ computations. $\chi^{2}$\ values are shown for the range of adopted
FeII shifts. In order to estimate confidence intervals we considered a set of fits with
displacements  $\Delta$\vr = +$500 n$, for integer $n = 0 ... 4$,
along with the best fit and a few additional $\Delta$\vr\  cases in proximity to the
minimum $\chi^{2}$. One can see a clear preference for zero or near-zero fits.
The significance of $\chi^{2}$ variations is described by $F$\ statistics appropriate for ratios
of $\chi^{2}$\ values \citep{bevington69}. Given the large number of degrees of freedoms
in the sampling range (4500 -- 4630, 5040  -- 5090, 5310 -- 5360 \AA) any $\chi^{2}$\ differences  between  two fits become significant at
a 95\%\ confidence level if $\chi^{2}/\chi_\mathrm{min}^{2} \approx 1.24$. The $\chi^{2}$\  values indicate that zero shift and ``best shift'' values in table 1 are not significantly different.
All fits involving shifts $\gtsim$ 500 \kms\ are statistically significant. The middle panel of
figure 1 upper row demonstrates visually that the fit with $\Delta $ \vr = +$1500 $ \kms\ (and even more
the fits with larger displacement) do not reproduce the observed FeII emission.

Both the residuals and $\chi^2$\ results rule out any systematic redshift for at least half of the H08
sample (pop. A).  Note especially the fits to the two relatively isolated multiplet 42 FeII lines
between H$\beta$ and [OIII]\l 4959 and on the red wing of [OIII]\l 5007. The redshifted fit fails
to include the blue side of the 4450 -- 4700 blend and the red side is confused  by the frequent
presence of HeII \l4686. The latter line is not mentioned in the H08  study leaving us to conclude that
it was not included in their fits. It can certainly give the impression of a redshift of the FeII blue blend, which is the most useful FeII diagnostic in the optical spectra of low redshift quasars (the red FeII blend is frequently affected by coronal lines as well as MgII host galaxy absorption in lower redshift sources).
Pioneering principal component analysis of the BQS survey \citep{borosongreen92} found that HeII\l4686 equivalent width anticorrelates with sources luminosity (it is Eigenvector 2). There is a tendency for the H08 sources with largest FeII redshifts to favor a smaller and lower ($\log L \sim 40.3 -  41.0$) range of source luminosity than those with near zero shifts ($\log L \sim 40.3 - 42.0$). Thus the effect of HeII will tend to play a larger role in the sources where the largest FeII redshifts have been found. H08 show composite spectra for five bins of FeII redshift in their Figure 12. The three bins involving largest FeII redshift  sources show a prominent HeII signature that, if not subtracted, will increase the apparent  significance of any assumed FeII redshift. Only the bin involving sources with no FeII redshift (within the uncertainties) shows no evidence of HeII emission.

\subsection{What about Population B Sources?}

Figure 9a of H08 suggests that a larger fraction of quasars with FWHM H$\beta >$ 4000 \kms (Population B) show large FeII redshifts. The lower panels of our Figure 1 show {\sc specfit} models superimposed on a B1 median
composite constructed from all H08 sources with quoted FeII redshift greater than 1000 \kms. The situation
is certainly more ambiguous than for the A2 composite. It is hard to identify individual FeII features. Lines are broad, FeII is weak and under these conditions there are serious limitations on the reliability of FWHM and
shift estimates for FeII (cf. Fig. 3 of \citealt{marzianietal03a}). The same analysis as done for A2 composite shows much poorer constraints on the FeII shift. The best fit yields $\Delta $ \vr $\approx 760$ \kms\ but is not distinguishable from a zero shift solution. If one actually computes $\chi^{2}$\ values over the ranges 4474 -- 4640 \AA,  5040 -- 5105 \AA, 5320 -- 5400 \AA, the $\chi^{2}$\  monotonically increases from 0 shift (figure 1, lower rightmost panel), although the increase remains  insignificant until $\approx$ 1100 \kms, where $\chi^{2}/\chi_\mathrm{min}^{2} \approx 1.21$. B1 FeII is too faint and the lines are too broad  to make inferences about line shifts and widths. The claim of large FeII shifts are not, and cannot be, confirmed.

\citet{kovacevicetal10} recently report an FeII
study of SDSS quasars and any FeII redshifts they measure (their figure 16) are much
smaller than those reported by H08 (the average FeII shift relative to the narrow lines is 100 $\pm$ 240 \kms).

Returning to our previous list of major sources of uncertainty
for FeII shift and FWHM estimates leads us to suggest that low
spectral s/n and above average HeII strength are the culprits. The fit to
the 4430 -- 4680 blue blend drives the best fit $\chi^2$ results. The
exclusion of HeII\l4686 from the H08 fits likely results in a tendency for
HeII to ``redshift'' the blue FeII blend. This effect in a typically low
luminosity sample, where HeII is stronger than average, likely drove the
conclusion that FeII was systematically redshifted. We tested this
conclusion omitting the HeII line from our fits to the bin A2 and B1
composites generated from the H08 sample. FeII shifts in lines 2 and 5
of table 1 increase from -60 to +770km/s and from 730 to 1570km/s, respectively. The more constraining A2 results
suggest that HeII can produce the entire systematic redshift claimed by H08.

\section{Conclusions}

We {\em do  not} confirm large FeII redshifts relative to narrow [OIII] and broad H$\beta$\ emission in Type 1 AGN but cannot rule out the existence of small red (or blue) shifts in particular subsamples. Fitting median composites built from spectra with large claimed FeII shifts ($\ge 1000$ \kms) indicates small shifts with an upper limit $\approx$ 300 \kms\ for bin A2. In the case of B1 the best fit suggests $\approx$700 \kms\ but the shift is very poorly constrained. In both cases the shifts are not significantly different from 0.  These results do not support the origin of FeII emission from a dynamical disjoint region from the one emitting the broad core of \hb. Our result also challenges the usefulness of FeII shift as orientation parameter. Small systematic shifts of FeII with respect to the rest frame seem plausible but a reliable analysis is possible only on spectra of high s/n ratio.

\begin{deluxetable}{lcccccc}
\tabletypesize{\scriptsize}
\setlength{\tabcolsep}{1pt}
%\rotate
\tablecaption{FeII and \hb\ shifts\tablenotemark{a} \label{tab:obs}}
\tablewidth{9.2cm}
\tablehead{
\colhead{Spectrum} & \colhead{$s \pm \delta s$(FeII)} & \colhead{$s$(\hb)\tablenotemark{b}}
& \colhead{$s$(\hb)}  & \colhead{~$|\Delta|$\tablenotemark{c}} &  \colhead{~~N spectra}\\
\colhead{}& \colhead{} &\colhead{BC} &\colhead{core} & \colhead{} & \colhead{} & \colhead{}
}
\startdata
A2 - Z10                    & -40 $\pm$ 90 &  10 &  0 &$\ltsim 20$  & 130\\
A2 - H08, $s \ge 1000$ &  -60 $\pm$ 400    & 80 & 90  &70  &  156\\

A2 - H08, $\eta \le 0.1$ & -160 $\pm$ 375 & 100 & 10 & 45 &  194\\

B1 - Z10  & -340 $\pm$ 400    & -150 & -80 &$\ltsim 20$  &  131\\
B1 - H08, $s \ge 1000$ &   +730$^{+400}_{-800}$ & 0 & 100& $\ltsim 20$ & 240\\
B1 - H08, $\eta \le 0.1$ &  +180 $\pm$ 450 &  -80  & 40& $\ltsim 20$ &  410\\
B1 - H08 $\cap$ Z10, $\eta \le 0.1$ & -150 $\pm$ 470 & -170 & -140 & 45 &  22\\

\enddata
\tablenotetext{a}{All shifts in \kms\ with respect to average radial velocity of peaks of \hb\ narrow component and of [OIII]\l 5007.}
\tablenotetext{b}{2$\sigma$ uncertainties of the H$\beta$\ BC measurements are
$\pm$60 \kms\ and $\pm$100 \kms\ for pop. A and B respectively.}
\tablenotetext{c}{Absolute value of peak radial velocity difference H$\beta$-[OIII]; $\Delta$\  is always positive if  larger than 20 \kms.  }

%\tablenotetext{b}{$\eta$ is the Eddington ratio.}
%\tablenotetext{c}{}
\tablecomments{Median composites listed in the first column: A2 - Z10, B1 - Z10: A2 and B1 sources in the Z10 sample; A2 and B1 - H08 $s \ge 1000$: H08 A2 and B1 sources with reported FeII shift $s \ge$ 1000\kms; A2 and B1 - H08 $\eta \le 0.1$: H08 A2 and B1 sources with Eddington ration $\eta$ $\le 0.1$\ (as estimated by H08). The last column indicates the numbers of spectra combined in the median composites. Z10-based composites have s/n $>$100, while lowest s/n H08-based composites have s/n $>$ 50. For the H08 subsamples we used spectra from the SDSS quasar catalog of \citet{shenetal11}}.

\end{deluxetable}

\bigskip

\begin{figure*}
\includegraphics[scale=0.26]{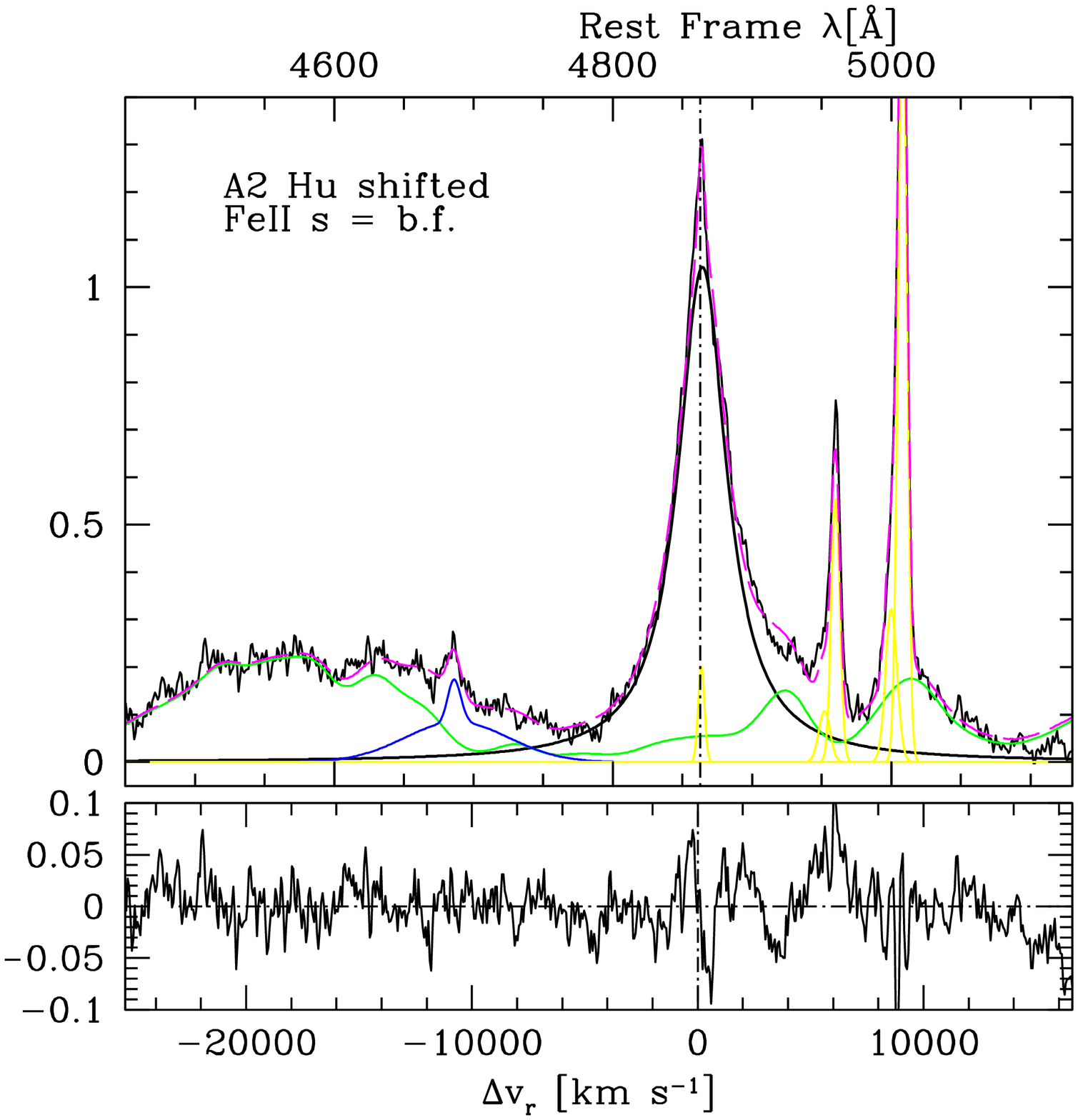}
\includegraphics[scale=0.26]{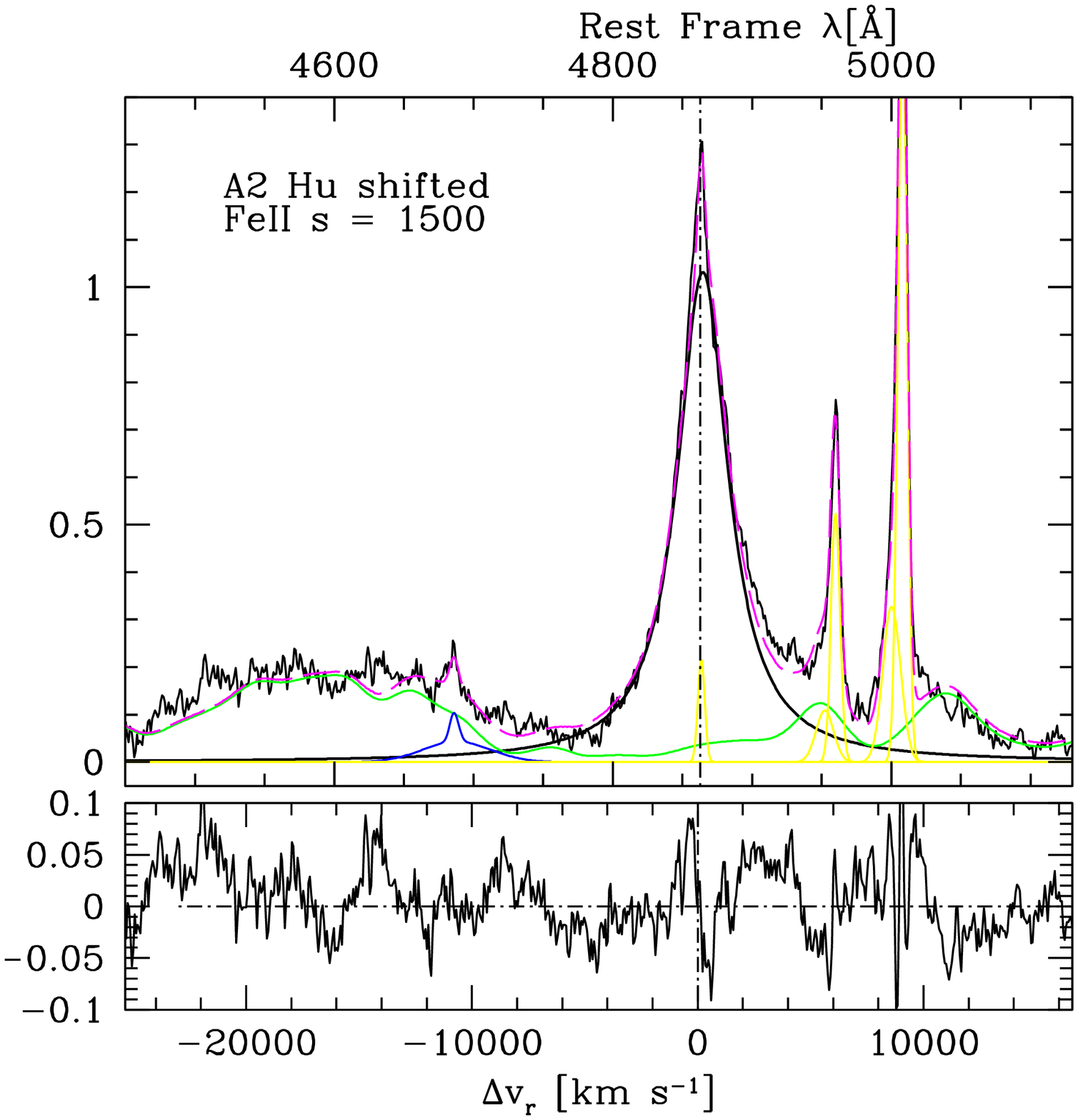}
\includegraphics[scale=0.26]{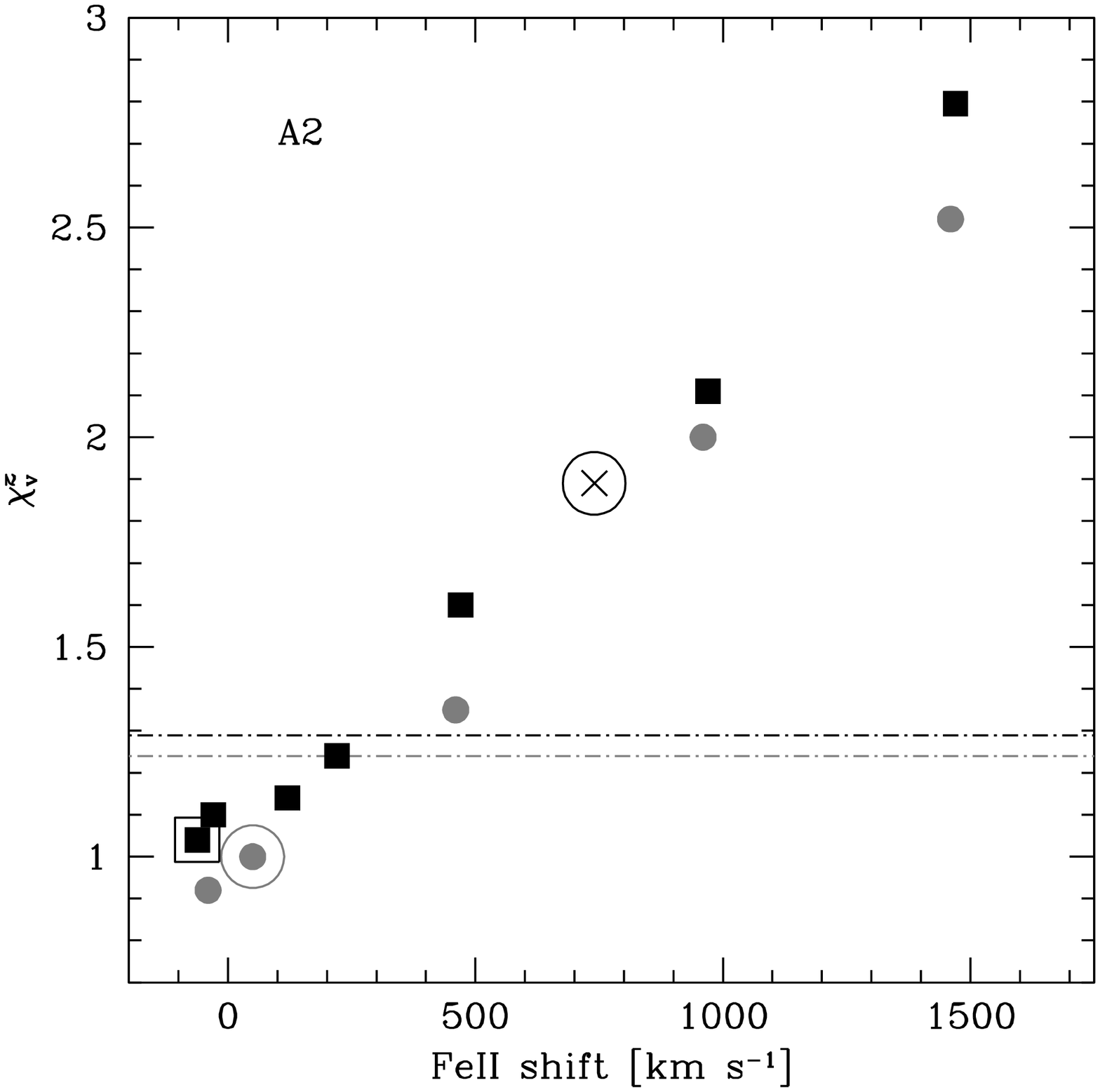}\\
\includegraphics[scale=0.26]{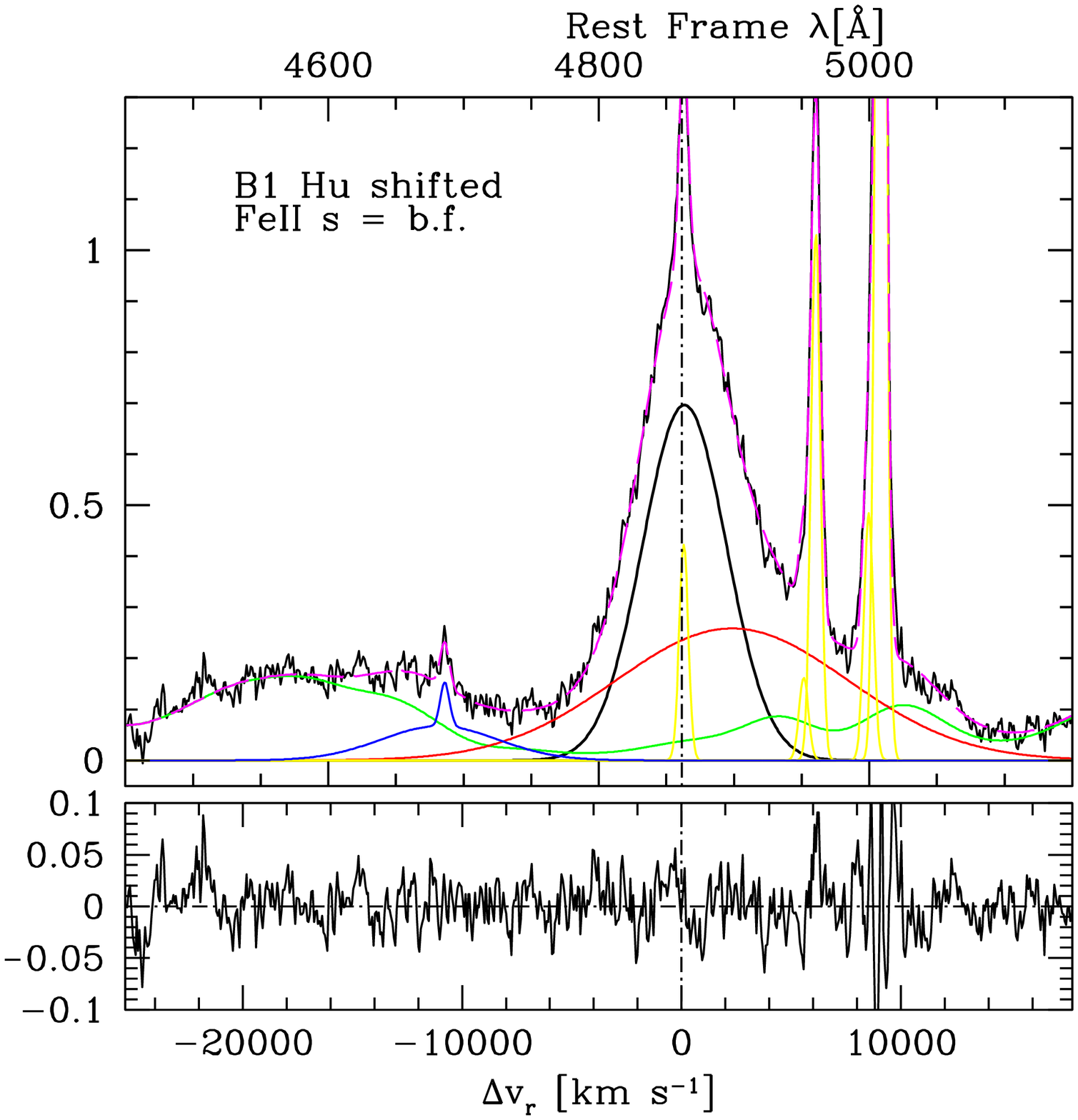}
\includegraphics[scale=0.26]{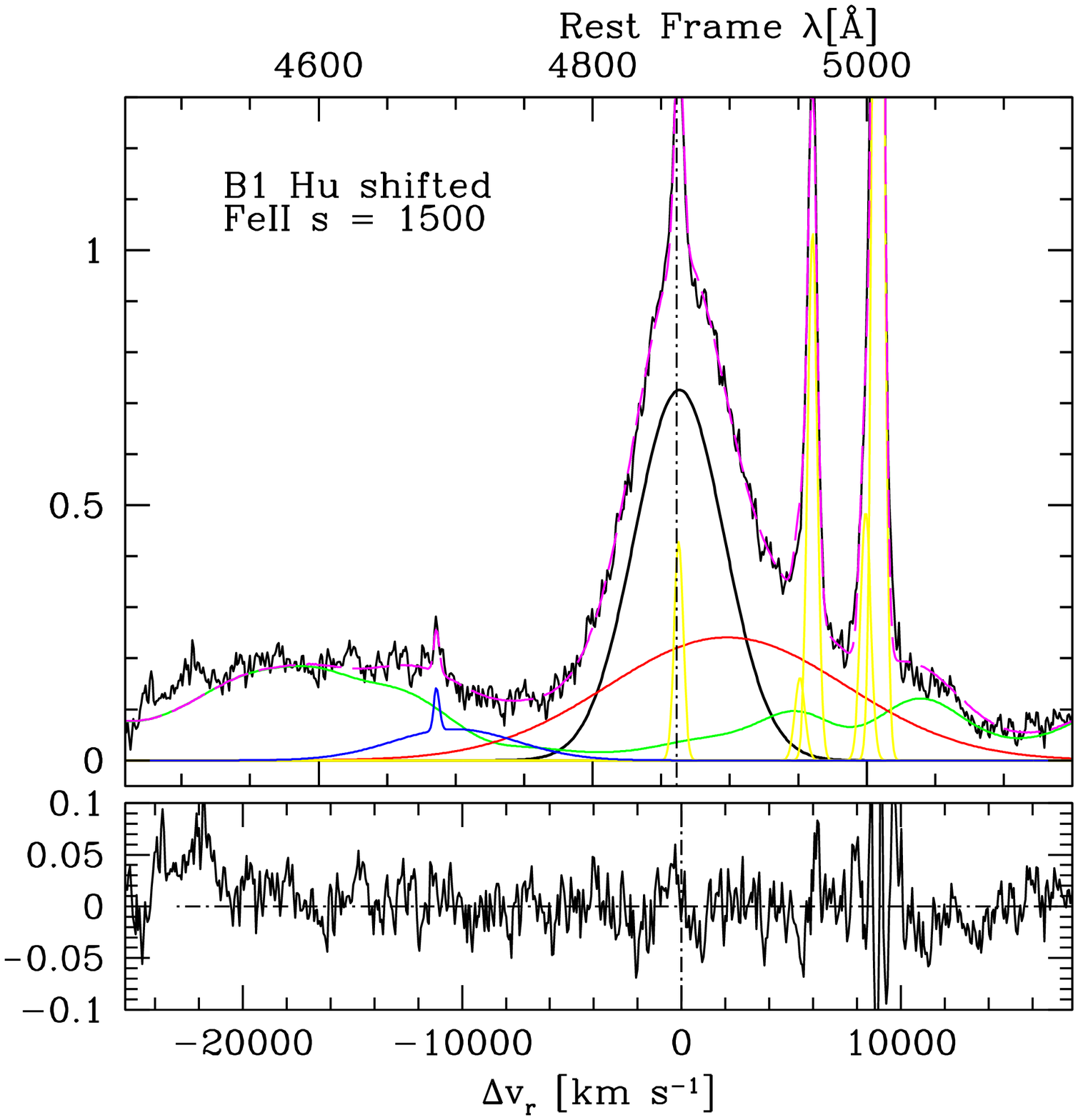}
\includegraphics[scale=0.26]{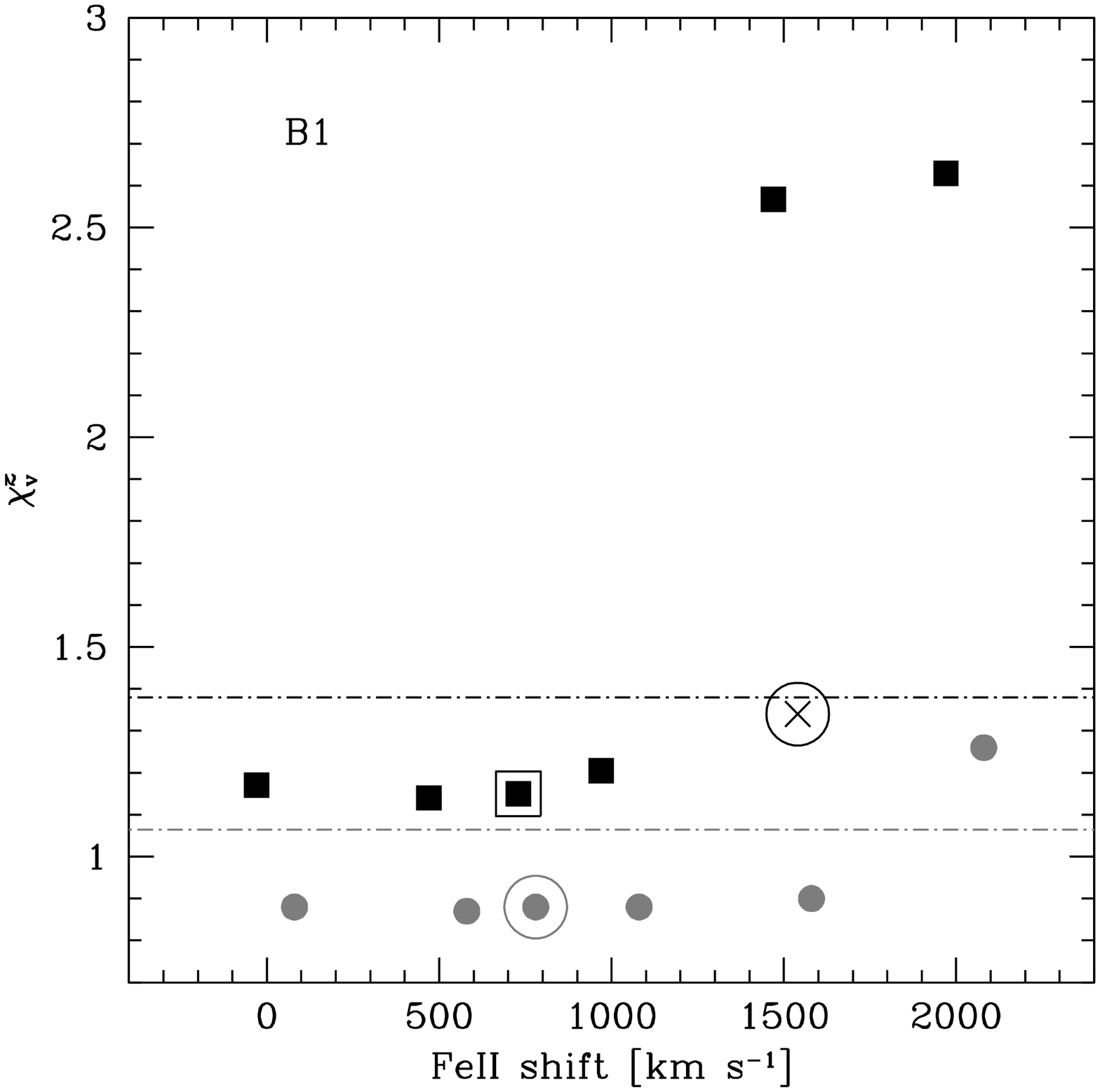}
\caption{Fits of the \hb\ spectral range (left and middle panels) and normalized $\chi^{2}_{\nu}$\ (right panels) as a function of  $\Delta$\vr\ of the FeII template. Top row: A2 median composite of spectra with reported FeII shift $\ge 1000$ \kms; bottom row: same for a composite of spectral type B1. Thick black line: broad component of \hb; thick red line: very broad component of \hb; green line: FeII template; yellow lines: narrow lines (\oiii\ and \hb\ narrow component). On the right panel, black squares and gray spots indicate the normalized $\chi^{2}_{\nu}$\ computed
%over the range 4500--4630 \AA\
on the composite with no restriction on FeII width, and with FeII width restricted to roughly match the distribution of H$\beta$\ width. Dot-dashed lines trace the minimum $\chi^{2}_{\nu}$ ratio indicating a significant difference from the best fit case. Circled symbols identify the {\sc specfit} best fit. The circled cross marks the best fit assuming no HeII \l 4686 contribution. Composites with no FeII width restriction have higher s/n making it possible to obtain more stringent limits on the maximum FeII shift.     \label{fig:contour}}
\end{figure*}


\begin{thebibliography}{27}
\expandafter\ifx\csname natexlab\endcsname\relax\def\natexlab#1{#1}\fi

\bibitem[{{Bachev} {et~al.}(2004){Bachev}, {Marziani}, {Sulentic}, {Zamanov},
  {Calvani}, \& {Dultzin-Hacyan}}]{bachevetal04}
{Bachev}, R., {Marziani}, P., {Sulentic}, J.~W., {Zamanov}, R., {Calvani}, M.,
  \& {Dultzin-Hacyan}, D. 2004, ApJ, 617, 171

\bibitem[{{Bevington}(1969)}]{bevington69}
{Bevington}, P.~R. 1969, {Data reduction and error analysis for the physical
  sciences} (New York: McGraw-Hill)

\bibitem[{{Boroson}(2005)}]{boroson05}
{Boroson}, T. 2005, \aj, 130, 381

\bibitem[{{Boroson}(2011)}]{boroson11}
{Boroson}, T.~A. 2011, \apjl, 735, L14+

\bibitem[{{Boroson} \& {Green}(1992)}]{borosongreen92}
{Boroson}, T.~A., \& {Green}, R.~F. 1992, ApJS, 80, 109

\bibitem[{{Ferland} {et~al.}(2009){Ferland}, {Hu}, {Wang}, {Baldwin}, {Porter},
  {van Hoof}, \& {Williams}}]{ferlandetal09}
{Ferland}, G.~J., {Hu}, C., {Wang}, J., {Baldwin}, J.~A., {Porter}, R.~L., {van
  Hoof}, P.~A.~M., \& {Williams}, R.~J.~R. 2009, \apjl, 707, L82

\bibitem[{{Gaskell}(1982)}]{gaskell82}
{Gaskell}, C.~M. 1982, ApJ, 263, 79

\bibitem[{{Gaskell}(1985)}]{gaskell85}
---. 1985, \nat, 315, 386

\bibitem[{{Hu} {et~al.}(2008){Hu}, {Wang}, {Ho}, {Chen}, {Bian}, \&
  {Xue}}]{huetal08}
{Hu}, C., {Wang}, J.-M., {Ho}, L.~C., {Chen}, Y.-M., {Bian}, W.-H., \& {Xue},
  S.-J. 2008, ApJ, 687, 78 (H08)

\bibitem[{{Kova{\v c}evi{\'c}} {et~al.}(2010){Kova{\v c}evi{\'c}},
  {Popovi{\'c}}, \& {Dimitrijevi{\'c}}}]{kovacevicetal10}
{Kova{\v c}evi{\'c}}, J., {Popovi{\'c}}, L.~{\v C}., \& {Dimitrijevi{\'c}},
  M.~S. 2010, \apjs, 189, 15

\bibitem[{{Kriss}(1994)}]{kriss94}
{Kriss}, G. 1994, Astronomical Data Analysis Software and Systems III, A.S.P.
  Conference Series, 61, 437

\bibitem[{{Marziani} {et~al.}(1993){Marziani}, {Sulentic}, {Calvani}, {Perez},
  {Moles}, \& {Penston}}]{Marzianietal93}
{Marziani}, P., {Sulentic}, J.~W., {Calvani}, M., {Perez}, E., {Moles}, M., \&
  {Penston}, M.~V. 1993, ApJ, 410, 56

\bibitem[{{Marziani} {et~al.}(2010){Marziani}, {Sulentic}, {Negrete},
  {Dultzin}, {Zamfir}, \& {Bachev}}]{Marzianietal10}
{Marziani}, P., {Sulentic}, J.~W., {Negrete}, C.~A., {Dultzin}, D., {Zamfir},
  S., \& {Bachev}, R. 2010, \mnras, 409, 1033

\bibitem[{{Marziani} {et~al.}(2009){Marziani}, {Sulentic}, {Stirpe}, {Zamfir},
  \& {Calvani}}]{marzianietal09}
{Marziani}, P., {Sulentic}, J.~W., {Stirpe}, G.~M., {Zamfir}, S., \& {Calvani},
  M. 2009, A\&Ap, 495, 83

\bibitem[{{Marziani} {et~al.}(2003{\natexlab{a}}){Marziani}, {Sulentic},
  {Zamanov}, {Calvani}, {Dultzin-Hacyan}, {Bachev}, \&
  {Zwitter}}]{marzianietal03a}
{Marziani}, P., {Sulentic}, J.~W., {Zamanov}, R., {Calvani}, M.,
  {Dultzin-Hacyan}, D., {Bachev}, R., \& {Zwitter}, T. 2003{\natexlab{a}},
  ApJS, 145, 199

\bibitem[{{Marziani} {et~al.}(2003{\natexlab{b}}){Marziani}, {Zamanov},
  {Sulentic}, \& {Calvani}}]{marzianietal03b}
{Marziani}, P., {Zamanov}, R.~K., {Sulentic}, J.~W., \& {Calvani}, M.
  2003{\natexlab{b}}, MNRAS, 345, 1133

\bibitem[{{Richards} {et~al.}(2011){Richards}, {Kruczek}, {Gallagher}, {Hall},
  {Hewett}, {Leighly}, {Deo}, {Kratzer}, \& {Shen}}]{richardsetal11}
{Richards}, G.~T., {et~al.} 2011, \aj, 141, 167

\bibitem[{{Shen} {et~al.}(2011)}]{shenetal11}
{Shen}, Y. {et~al.} 2011, \apjs, 194, 45

\bibitem[{{Shields} {et~al.}(2010)}]{shieldsetal10}{Shields}. G., {et~al} 2010, \apj, 687, 78

\bibitem[{{Sulentic} {et~al.}(2007){Sulentic}, {Bachev}, {Marziani}, {Negrete},
  \& {Dultzin}}]{sulenticetal07}
{Sulentic}, J.~W., {Bachev}, R., {Marziani}, P., {Negrete}, C.~A., \&
  {Dultzin}, D. 2007, ApJ, 666, 757

\bibitem[{{Sulentic} {et~al.}(2000{\natexlab{a}}){Sulentic}, {Marziani}, \&
  {Dultzin-Hacyan}}]{sulenticetal00a}
{Sulentic}, J.~W., {Marziani}, P., \& {Dultzin-Hacyan}, D. 2000{\natexlab{a}},
  ARA\&A, 38, 521

\bibitem[{{Sulentic} {et~al.}(2002){Sulentic}, {Marziani}, {Zamanov}, {Bachev},
  {Calvani}, \& {Dultzin-Hacyan}}]{sulenticetal02}
{Sulentic}, J.~W., {Marziani}, P., {Zamanov}, R., {Bachev}, R., {Calvani}, M.,
  \& {Dultzin-Hacyan}, D. 2002, ApJL, 566, L71

\bibitem[{{Sulentic} {et~al.}(2000{\natexlab{b}}){Sulentic}, {Marziani},
  {Zwitter}, {Dultzin-Hacyan}, \& {Calvani}}]{sulenticetal00b}
{Sulentic}, J.~W., {Marziani}, P., {Zwitter}, T., {Dultzin-Hacyan}, D., \&
  {Calvani}, M. 2000{\natexlab{b}}, ApJL, 545, L15

\bibitem[{{Sulentic} {et~al.}(2006){Sulentic}, {Repetto}, {Stirpe}, {Marziani},
  {Dultzin-Hacyan}, \& {Calvani}}]{sulenticetal06}
{Sulentic}, J.~W., {Repetto}, P., {Stirpe}, G.~M., {Marziani}, P.,
  {Dultzin-Hacyan}, D., \& {Calvani}, M. 2006, A\&Ap, 456, 929

\bibitem[{{V{\'e}ron-Cetty} {et~al.}(2001){V{\'e}ron-Cetty}, {V{\'e}ron}, \&
  {Gon{\c c}alves}}]{veroncettyetal01}
{V{\'e}ron-Cetty}, M.-P., {V{\'e}ron}, P., \& {Gon{\c c}alves}, A.~C. 2001,
  AAp, 372, 730

\bibitem[{{Wang} {et~al.}(2005){Wang}, {Dong}, {Zhang}, {Zhou}, {Wang},
  \& {Lu}}]{wangetal05}
{Wang}, T. -G., {Dong}, X. -B., {Zhang}, X., -G., {Zhou}, H., -Y., {Wang}, J. -X. \&
  {Lu}, Y., -J. 2005, ApJL, 625, L35

\bibitem[{{Wang} {et~al.}(2011){Wang}, {Wang}, {Zhou}, {Liu}, {Wang}, {Yuan},
  \& {Dong}}]{wangetal11}
{Wang}, H., {Wang}, T., {Zhou}, H., {Liu}, B., {Wang}, J., {Yuan}, W., \&
  {Dong}, X. 2011, ArXiv e-prints

\bibitem[{{Zamanov} {et~al.}(2002){Zamanov}, {Marziani}, {Sulentic}, {Calvani},
  {Dultzin-Hacyan}, \& {Bachev}}]{zamanovetal02}
{Zamanov}, R., {Marziani}, P., {Sulentic}, J.~W., {Calvani}, M.,
  {Dultzin-Hacyan}, D., \& {Bachev}, R. 2002, ApJL, 576, L9

\bibitem[{{Zamfir} {et~al.}(2008){Zamfir}, {Sulentic}, \& {Marziani}}]{zamfiretal08}
{Zamfir}, S., {Sulentic}, J.~W., \& {Marziani}, P., 2008, \mnras,
  403, 1759

\bibitem[{{Zamfir} {et~al.}(2010){Zamfir}, {Sulentic}, {Marziani}, \&
  {Dultzin}}]{zamfiretal10}
{Zamfir}, S., {Sulentic}, J.~W., {Marziani}, P., \& {Dultzin}, D. 2010, \mnras,
  387, 856 (Z10)

\bibitem[{{Zhou} {et~al.}(2006){Zhou}, {Wang}, {Yuan}, {Lu}, {Dong}, {Wang}, \&
  {Lu}}]{zhouetal06}
{Zhou}, H., {Wang}, T., {Yuan}, W., {Lu}, H., {Dong}, X., {Wang}, J., \& {Lu},
  Y. 2006, ApJS, 166, 128

\end{thebibliography}
\end{document}